# A NEW MESHLESS "FRAGILE POINTS METHOD (FPM)" BASED ON A GALERKIN WEAK-FORM FOR 2D FLEXOELECTRIC ANALYSIS

Yue Guan[1], Leiting Dong[2], Satya N. Atluri[1]

[1]Texas Tech University, Lubbock, TX
[2]Beihang University, Beijing, China

**ABSTRACT**

*A meshless Fragile Points Method (FPM) is presented for analyzing 2D flexoelectric problems. Local, simple, polynomial and discontinuous trial and test functions are generated with the help of a local meshless differential quadrature approximation of the first three derivatives. Interior Penalty Numerical Fluxes are employed to ensure the consistency of the method. Based on a Galerkin weak-form formulation, the present FPM leads to symmetric and sparse matrices, and avoids the difficulties of numerical integration in the previous meshfree methods. Numerical examples including isotropic and anisotropic materials with flexoelectric and piezoelectric effects are provided as validations. The present method is much simpler than the Finite Element Method, or the Element-Free Galerkin (EFG) and Meshless Local Petrov-Galerkin (MLPG) methods, and the numerical integration of the weak form is trivially simple.*

## 1. INTRODUCTION

The flexoelectricity describes an electromechanical coupling effect between the electric polarization and mechanical strain gradient [1-3]. As a result of the recent miniaturized trend of electromechanical systems, which leads a tremendous growth in the scale of strain gradient, there has been an increasing demand for reliable and accurate theories and numerical methods for flexoelectric analysis [2,4]. In 2006, Maranganti et al. [5] proposed the first continuum theory for flexoelectricity, in which the internal energy density is given as a function of strain, strain gradient, electric polarization and polarization gradient. After that, a number of advanced continuum theories have been carried out, considering the Maxwell stress [6], the nonlinear flexoelectric behavior under larger deformation [7], etc.

Generally, the flexoelectric behavior is governed by a fourth-order partial differential equation. As a result, the main difficulty in modelling flexoelectricity for classic element-based methods lies in the $C^1$ continuity requirement. On one hand, the Finite Element Method (FEM) based on Argyris triangular element [8] can be employed to remedy the $C^1$ requirement. Sladek et al. [9] have also proposed a $C^1$ continuous FEM analyzing the effects of electric field and strain gradients. However, both of the approaches require 9 degrees of freedom (DOFs) at each node, and thus make it computationally expensive. Alternatively, some mixed FEM [3,4] using Lagrangian multipliers as independent variables are also available, requiring only $C^0$ continuity. Yet they have even more DOFs in each element, making it prohibitively expensive for practical use.

On the other hand, meshless methods like the Local Maximum-Entropy (LME) Meshfree Method [10], Element-Free Galerkin (EFG) Method [11], and Meshless Local Petrov-Galerkin (MLPG) Method [12] have also shown their capability in analyzing flexoelectric or strain-gradient effects. In the EFG and MLPG method, with the help of Moving Least Square (MLS) approximation, global shape and trial functions with $C^2$ continuity can be generated. However, these trial functions based on MLS are rational and considerably complicated. As a result, their numerical integration in the weak form in either EFG or MLPG can be tedious. Actually, the difficulty in domain integration is a common problem in many Galerkin meshfree methods [13]. The simplest choice is direct nodal integration. Though efficient, and no background mesh required, the direct nodal integration may have stability issues and can be less accurate or non-convergent. A number of modified nodal integration methods are proposed, in order to ensure the accuracy and stability, which in turn sacrifice the efficiency [13]. Some other newer-type quadrature schemes are also adopted [14].

Alternatively, the difficulty of numerical integration can be solved fundamentally by employing simple shape functions like polynomials. The Fragile Points Method (FPM), in contrast to most of the previous meshless methods, is based on simple, polynomial, piecewise-continuous trial and test functions. Consequently, the classic Gaussian quadrature is applied, and numerical integrals are relatively simple, especially when simple subdomain shapes are used (e.g., quadrilateral or triangle). Numerical Flux Corrections are employed to ensure the continuity condition. On the other hand, when compared with the element-based methods, the FPM is generated using point stiffness matrices. Thus, it has benefits in inserting or removing points and bypassing the mesh distortion problem. The discontinuity of the trial function also presents a great potential in analyzing systems involving cracks, ruptures and fragmentations. Furthermore, when compared with the EFG and MLPG methods, the FPM also has advantages like symmetric matrices, and the shape functions pass through the nodal data directly.

The FPM has already shown great accuracy and efficiency in solving 2D thermal conductivity and elasticity problems. In this study, we formulate and apply the FPM for analyzing linear flexoelectric problems.

## 2. FLEXOELECTRICITY THEORY AND BOUNDARY-VALUE PROBLEM

Phenomenologically, the flexoelectricity describes an electric polarization generated by the mechanical strain-gradient:

$$P_i = \bar{\mu}_{ijkl}\kappa_{jkl}, \quad (1)$$

where $\mathbf{P} = [P_1, P_2]^T$ is the electrical polarization vector. $\mathbf{u} = [u_1, u_2]^T$ is the displacement vector in a 2D domain $\Omega$. $\kappa_{jkl} = u_{l,jk}$ is the second gradient of displacement (i.e., the strain-gradient), and $\bar{\mu}_{ijkl}$ are elements of a fourth order flexoelectric tensor ($i,j,k,l = 1,2$). Einstein summation convention is used here.

In the current study, the linear flexoelectric theory proposed by [5] is employed, in which the internal energy is described as a function of strain $\varepsilon_{ij}$, strain gradient $\kappa_{jki}$ and polarization $P_i$, i.e. $\tilde{U} = \tilde{U}(\boldsymbol{\varepsilon}, \boldsymbol{\kappa}, \mathbf{P})$. The corresponding governing equations can be written as:

$$\left(\sigma_{ij} - \mu_{jki,k}\right)_{,j} + b_i = 0, \quad (2)$$
$$-\epsilon_0 \phi_{,ii} + P_{i,i} = q \quad (3)$$

where $\boldsymbol{\sigma}$ is the Cauchy stress, $\boldsymbol{\mu}$ is the double-stress (conjugate of the strain gradient $\boldsymbol{\kappa}$), $\phi$ is the electrical potential, $\mathbf{b}$ is the body force per volume, $q$ is the free charge per volume, and $\epsilon_0$ is the permittivity of free space. The corresponding boundary conditions are

$$u_i = \tilde{u}_i, \quad \text{on } \partial\Omega_u, \quad (4)$$
$$\left(\sigma_{ij} - \mu_{jki,k}\right)n_j + \left[(\nabla_p^t n_p)n_k - \nabla_k^t\right](n_m \mu_{mki})$$
$$= \tilde{Q}_i, \quad \text{on } \partial\Omega_Q, \quad (5)$$
$$\nabla^n u_i = \tilde{d}_i, \quad \text{on } \partial\Omega_d, \quad (6)$$
$$n_j n_k \mu_{jki} = \tilde{R}_i, \quad \text{on } \partial\Omega_R, \quad (7)$$
$$\phi = \tilde{\phi}, \quad \text{on } \partial\Omega_\phi, \quad (8)$$
$$\left(-\epsilon_0 \phi_{,i} + P_i\right)n_i = -\tilde{\omega}, \quad \text{on } \partial\Omega_\omega, \quad (9)$$

where $(\tilde{u}, \tilde{Q}, \tilde{d}, \tilde{R}, \tilde{\phi}, \tilde{\omega})$ are known functions, $\nabla^n = \mathbf{n} \cdot \nabla$ is the normal derivative, $\nabla^t = \nabla - \mathbf{n}\nabla^n$ the 'surface gradient' on $\partial\Omega$. $\partial\Omega_u \cup \partial\Omega_Q = \partial\Omega_d \cup \partial\Omega_R = \partial\Omega_\phi \cup \partial\Omega_\omega = \partial\Omega$, and $\partial\Omega_u \cap \partial\Omega_Q = \partial\Omega_d \cap \partial\Omega_R = \partial\Omega_\phi \cap \partial\Omega_\omega = \emptyset$.

We consider a material with both piezoelectric and flexoelectric effects. The constitutive equations are:

$$\boldsymbol{\sigma} = \mathbf{D}_{\sigma\varepsilon}\boldsymbol{\varepsilon} - \mathbf{G}_0\boldsymbol{\kappa} - \mathbf{e}\mathbf{E}, \quad (10)$$
$$\boldsymbol{\mu} = \mathbf{D}_{\mu\kappa}\boldsymbol{\kappa} - \mathbf{G}_0^T\boldsymbol{\varepsilon} - \mathbf{A}_0\mathbf{E} \quad (11)$$
$$\mathbf{P} = (\bar{\boldsymbol{\kappa}} - \epsilon_0 \mathbf{I})\mathbf{E} + \mathbf{e}^T\boldsymbol{\varepsilon} + \mathbf{A}_0^T\boldsymbol{\kappa}, \quad (12)$$

where

$$\boldsymbol{\varepsilon} = [\varepsilon_{11}, \varepsilon_{22}, 2\varepsilon_{12}]^T, \boldsymbol{\sigma} = [\sigma_{11}, \sigma_{22}, \sigma_{12}]^T,$$
$$\boldsymbol{\kappa} = [\kappa_{111}, \kappa_{222}, 2\kappa_{121}, 2\kappa_{122}, \kappa_{221}, \kappa_{112}]^T,$$
$$\boldsymbol{\mu} = [\mu_{111}, \mu_{222}, \mu_{121}, \mu_{122}, \mu_{221}, \mu_{112}]^T,$$
$$\mathbf{E} = [E_1, E_2]^T,$$

and

$$\varepsilon_{ij} = \frac{1}{2}(u_{i,j} + u_{j,i}), \quad \sigma_{ij} = \sigma_{ji},$$
$$\kappa_{kji} = \kappa_{jki}, \qquad \mu_{kji} = \mu_{jki},$$
$$E_i = -\phi_{,i}.$$

The constitutive matrices:

$$\mathbf{D}_{\sigma\varepsilon} = \bar{\mathbf{D}}_{\sigma\varepsilon} - \mathbf{e}(\bar{\boldsymbol{\kappa}} - \epsilon_0 \mathbf{I})^{-1}\mathbf{e}^T, \quad (13)$$
$$\mathbf{D}_{\mu\kappa} = \bar{\mathbf{D}}_{\mu\kappa} - \mathbf{A}_0(\bar{\boldsymbol{\kappa}} - \epsilon_0 \mathbf{I})^{-1}\mathbf{A}_0^T, \quad (14)$$
$$\mathbf{G}_0 = \mathbf{e}(\bar{\boldsymbol{\kappa}} - \epsilon_0 \mathbf{I})^{-1}\mathbf{A}_0^T. \quad (15)$$

$\bar{\mathbf{D}}_{\sigma\varepsilon}$, $\bar{\mathbf{D}}_{\mu\kappa}$, $\bar{\boldsymbol{\kappa}}$, $\mathbf{A}_0$ and $\mathbf{e}$ are matrices of material properties (see Appendix A). $\mathbf{I}$ is a unit matrix. In the numerical calculations, for simplicity, we only concentrate on materials with isotropic elasticity, cubic symmetry for the flexoelectric tensor, and tetragonal symmetry for the piezoelectric tensor.

## 3. TRIAL AND TEST FUNCTIONS

### 3.1 Points and domain partition

First, a set of random points are scattered in the domain. The entire domain can then be partitioned into several nonoverlapping subdomains. Within each subdomain, only one point exists. Multiple partitions are possible here, e.g., the Voronoi Diagram partition (see Fig. 1(a)), quadrilateral partition (see Fig. 1(b)), and triangular partition. The traditional FEM meshing can also be employed. The elements can be converted into subdomains, while the internal point is defined as the center of mass of each subdomain. Thus, the preprocessing module of commercial FEA software like ABAQUS can be helpful for generating the points and subdomains in this proposed FPM.

However, unlike the FEM, the trial and test functions in the present FPM are *point-based*. In order to describe the high-order behavior, the trial function $\mathbf{u}^h$ in each subdomain can be written in the form of a third-order Taylor expansion at the corresponding internal point. For instance, in subdomain $E_0$ with internal point $P_0$:

$$\mathbf{u}^h(x,y) = \mathbf{u}^0 + (x-x_0)\mathbf{u}_{,1}\big|_{P_0} + (y-y_0)\mathbf{u}_{,2}\big|_{P_0} +$$
$$\frac{1}{2}(x-x_0)^2 \mathbf{u}_{,11}\big|_{P_0} + (x-x_0)(y-y_0)\mathbf{u}_{,12}\big|_{P_0} +$$
$$\frac{1}{2}(y-y_0)^2 \mathbf{u}_{,22}\big|_{P_0} + \frac{1}{6}(x-x_0)^3 \mathbf{u}_{,111}\big|_{P_0} + \frac{1}{2}(x-$$

$$x_0)^2(y-y_0)\mathbf{u}_{,112}\big|_{P_0} + \frac{1}{2}(x-x_0)(y-y_0)^2\mathbf{u}_{,122}\big|_{P_0} + \frac{1}{6}(y-y_0)^3\mathbf{u}_{,222}\big|_{P_0}, \quad (16)$$

where $(x_0, y_0)$ are the coordinates of the point $P_0$. $\mathbf{u}^0$ is the value of $\mathbf{u}^h$ at $P_0$. The first three derivatives $[\overline{\mathbf{D}}\mathbf{u}]|_{P_0} = [\mathbf{u}_{,1}^T, \mathbf{u}_{,2}^T, \mathbf{u}_{,11}^T, \mathbf{u}_{,12}^T, \mathbf{u}_{,22}^T, \mathbf{u}_{,111}^T, \mathbf{u}_{,112}^T, \mathbf{u}_{,122}^T, \mathbf{u}_{,222}^T]^T\big|_{P_0}$ are yet unknown. Here we introduce the local differential quadrature method to determine the high-order derivatives.

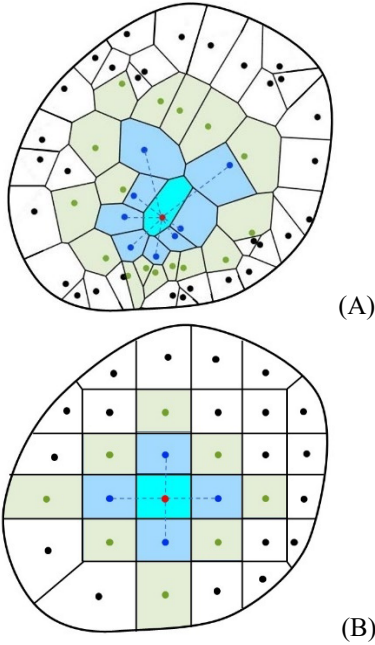

**FIGURE 1:** THE DOMAIN $\Omega$ AND ITS PARTITIONS. (A) VORONOI DIAGRAM PARTITION. (B) QUADRILATERAL PARTITION.

### 3.2 Local Radial Basis Function-based Differential Quadrature method

The conventional differential quadrature (DQ) method is a widely used numerical discretization technique [15]. However, usually based on 1D basis functions (e.g., polynomials), the conventional DQ method can only be applied along a mesh-line and cannot be incorporated directly with meshless schemes. In recent years, several multi-dimensional DQ methods have been developed based on the meshless Radial Basis Functions (RBFs). In the current work, we employ the local RBF-DQ method proposed by Shu et al. [15] in 2003 which approximates the derivatives directly using a limited number of supporting points.

In the current methodology, as shown in Fig. 1, for a given point $P_0 \in E_0$ (red), its support is defined to involve all its nearest (blue) and second (green) neighboring points. Here the nearest neighboring points are the points in subdomains sharing boundaries with $E_0$, while the second neighboring points are nearest neighbors of the former. For the points on or close to the domain boundary $\partial\Omega$, the third neighboring points are also taken into consideration to remedy the lack of effective supporting points. These supporting points are named as $P_1, P_2, \dots, P_m$.

Many RBFs can be used as basis function in the local RBF-DQ method. The multiquadric (MQ), inverse-MQ and Gaussian RBFs are all available in conjunction with the FPM and achieve similar accuracy with appropriate parameters. Here we use the MQ-RBF as an example:

$$f(r) = \sqrt{r^2 + c^2}, \quad (17)$$

where $r$ is the radial from the conference point. $c = c_0 D_0$ is a constant parameter, where $D_0$ is the diameter of the minimal circle enclosing all the supporting points.

Thus, any partial derivative of displacement at point $P_0$ can be approximated by a weighted linear sum of the value of $\mathbf{u}^h$ at all its supporting points:

$$\frac{\partial^{s+t}}{\partial x^s \partial y^t}\mathbf{u}^h\bigg|_{P_0} = \sum_{i=0}^{m} W_i^{(s,t)} \mathbf{u}^h|_{P_i}, \quad (18)$$

where $W_i^{(s,t)}$ is the weighting coefficient corresponding to point $P_i$. After rearrangement, the derivatives under study at point $P_0$ can be approximated as:

$$[\overline{\mathbf{D}}\mathbf{u}]|_{P_0} = (\overline{\mathbf{B}} \otimes \mathbf{I}_{2\times 2})\mathbf{u}_E, \quad (19)$$

where

$$\mathbf{u}_E = [u_1^0, u_2^0, u_1^1, u_2^1, \dots, u_1^m, u_2^m]^T. \quad (20)$$

$\mathbf{u}^i = [u_1^i, u_2^i]^T$ is the value of $\mathbf{u}^h$ at $P_i$. $\overline{\mathbf{B}}$ is the matrix of weighting coefficients $W_i^{(s,t)}$, $\mathbf{I}_{2\times 2}$ is a unit matrix, and $\otimes$ donates the Kronecker product. The weighting coefficient matrix $\overline{\mathbf{B}}$ can be obtained from:

$$\overline{\mathbf{B}} = \mathbf{G}^{-1}[\overline{\mathbf{D}}\mathbf{G}], \quad (21)$$

where

$$\mathbf{G} = \begin{bmatrix} 1 & 1 & \cdots & 1 \\ g_1(x_1, y_1) & g_1(x_2, y_2) & \cdots & g_1(x_m, y_m) \\ \vdots & \vdots & \ddots & \vdots \\ g_m(x_1, y_1) & g_m(x_2, y_2) & \cdots & g_m(x_m, y_m) \end{bmatrix},$$

$$[\overline{\mathbf{DG}}] = \begin{bmatrix} 0 & g_{1,1}(x_1,y_1) & \cdots & g_{m,1}(x_1,y_1) \\ 0 & g_{1,2}(x_1,y_1) & \cdots & g_{m,2}(x_1,y_1) \\ 0 & g_{1,11}(x_1,y_1) & \cdots & g_{m,11}(x_1,y_1) \\ 0 & g_{1,12}(x_1,y_1) & \cdots & g_{m,12}(x_1,y_1) \\ 0 & g_{1,22}(x_1,y_1) & \cdots & g_{m,22}(x_1,y_1) \\ 0 & g_{1,111}(x_1,y_1) & \cdots & g_{m,111}(x_1,y_1) \\ 0 & g_{1,112}(x_1,y_1) & \cdots & g_{m,112}(x_1,y_1) \\ 0 & g_{1,122}(x_1,y_1) & \cdots & g_{m,122}(x_1,y_1) \\ 0 & g_{1,222}(x_1,y_1) & \cdots & g_{m,222}(x_1,y_1) \end{bmatrix}^{\mathrm{T}},$$

$$g_i(x,y) = \sqrt{(x-x_i)^2 + (y-y_i)^2 + c^2} - \sqrt{(x-x_0)^2 + (y-y_0)^2 + c^2}.$$

Note that the accuracy of local RBF-DQ approximation is excellent for the first derivative. Whereas for higher-order approximations, the accuracy decreases yet are still acceptable. This implies that a mixed formulation may help to improve the accuracy of the current method.

### 3.3 Local, polynomial, discontinuous test and trial functions

Substitute the approximation of derivatives into Eqn. (16). The trial function $\mathbf{u}^h$ in subdomain $E_0$ is achieved:

$$\mathbf{u}^h(x,y) = \mathbf{N}(x,y)\mathbf{u}_E, \tag{22}$$

where

$$\mathbf{N}(x,y) = \left(\overline{\mathbf{N}} \cdot \overline{\mathbf{B}} + \begin{bmatrix} 1 & 0 & \cdots & 0 \end{bmatrix}_{1\times(m+1)}\right) \otimes \mathbf{I}_{2\times 2},$$

$$\overline{\mathbf{N}} = \Big[x-x_0,\ y-y_0, \tfrac{1}{2}(x-x_0)^2,\ (x-x_0)(y-y_0),\ \tfrac{1}{2}(y-y_0)^2,\ \tfrac{1}{6}(x-x_0)^3,\ \tfrac{1}{2}(x-x_0)^2(y-y_0),\ \tfrac{1}{2}(x-x_0)(y-y_0)^2,\ \tfrac{1}{6}(y-y_0)^3\Big].$$

The trial function and shape function are defined following the same process in each subdomain. Since no continuity requirement is employed at the internal boundaries, the shape and trial functions can be discontinuous. Figure 2(A) shows the graph of a shape function in a unit domain with 50 random points and Voronoi Diagram partition. The trial function simulating an exponential function $u_a = e^{-10\sqrt{(x-0.5)^2+(y-0.5)^2}}$ is presented in Fig. 2(B). As can be seen, the trial function is a cubic polynomial in each subdomain. It is piecewise-continuous. However, as the shape function in each subdomain also depends on the values of the neighboring points, the entire shape function shows a weakly continuous tendency. The shape and trial functions for the electric potential $\phi^h$ are generated similarly.

The test function for displacement $\mathbf{v}^h$ and electrical potential $\tau^h$ in the Galerkin weak-form in the FPM are prescribed to possess the same shape as $\mathbf{u}^h$ and $\phi^h$ respectively. As a result of the discontinuous trial functions, the FPM also has a great potential in analyzing systems involving cracks, ruptures and fragmentations.

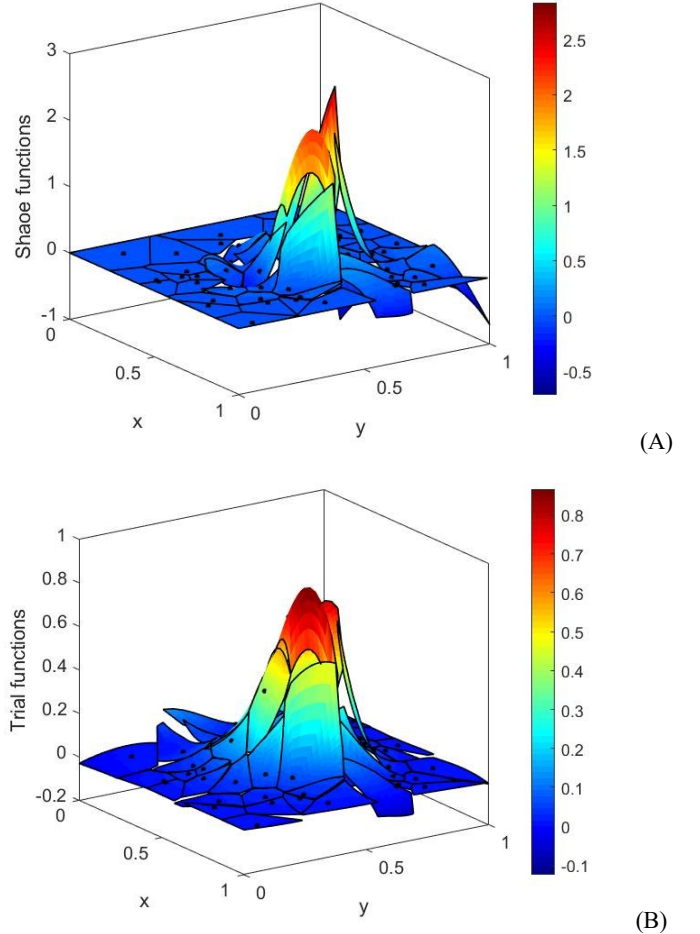

**FIGURE 2:** THE SHAPE FUNCTION AND TRIAL FUNCTION.

### 4. Weak-form formulation and Numerical Flux Corrections

We multiply the governing equations (2) and (3) by the test functions $v_i^h$ and $\tau^i$ respectively in each subdomain $E$, then apply the Gauss divergence theorem:

$$\begin{aligned} &\int_E v_{i,j}\sigma_{ij}\,\mathrm{d}\Omega + \int_E v_{i,jk}\sigma_{kji}\,\mathrm{d}\Omega \\ &= \int_E v_i\,b_i\,\mathrm{d}\Omega + \int_{\partial E} v_i(\sigma_{ij}-\mu_{kji,k})n_j\,\mathrm{d}\Gamma \\ &\quad + \int_{\partial E} v_{i,j}\mu_{kji}n_k\,\mathrm{d}\Gamma \end{aligned} \tag{23}$$

$$\int_E \epsilon_0 \tau_{,i} \phi_{,i} \, d\Omega - \int_E \tau_{,i} P_i \, d\Omega = \int_E \tau q \, d\Omega$$
$$- \int_{\partial E} \tau(-\epsilon_0 \phi_{,i} + P_i) n_i \, d\Gamma \tag{24}$$

where $\partial E$ is the boundary of the subdomain, $\mathbf{n} = [n_1, n_2]^T$ is the unit vector outward to $\partial E$.

Let $\Gamma_h$ donates the set of all internal boundaries, and $\Gamma = \Gamma_h + \partial \Omega = \Gamma_h + \partial \Omega_u + \partial \Omega_Q = \Gamma_h + \partial \Omega_d + \partial \Omega_R = \Gamma_h + \partial \Omega_\phi + \partial \Omega_\omega$. By summing the equations over the entire domain, we rewrite them in the matrix forms:

$$\int_\Omega \boldsymbol{\varepsilon}^T(\mathbf{v}) \boldsymbol{\sigma}(\mathbf{u}, \phi) \, d\Omega + \int_\Omega \boldsymbol{\kappa}^T(\mathbf{v}) \boldsymbol{\mu}(\mathbf{u}, \phi) \, d\Omega$$
$$- \int_{\Gamma_h} [\![\hat{\boldsymbol{\varepsilon}}^T(\mathbf{v})]\!] \bar{\mathbf{n}}_3^e \{\boldsymbol{\mu}(\mathbf{u}, \phi)\} \, d\Gamma$$
$$- \int_{\Gamma_h} \{\hat{\boldsymbol{\varepsilon}}^T(\mathbf{v})\} \bar{\mathbf{n}}_3^e [\![\boldsymbol{\mu}(\mathbf{u}, \phi)]\!] \, d\Gamma$$
$$- \int_{\Gamma_h} [\![\mathbf{v}^T]\!] \left\{ \begin{matrix} \bar{\mathbf{n}}_1^e \boldsymbol{\sigma}(\mathbf{u}, \phi) - \bar{\mathbf{n}}_{21}^e \boldsymbol{\mu}_{,1}(\mathbf{u}, \phi) \\ -\bar{\mathbf{n}}_{22}^e \boldsymbol{\mu}_{,2}(\mathbf{u}, \phi) \end{matrix} \right\} d\Gamma$$
$$- \int_{\Gamma_h} \{\mathbf{v}^T\} \left[\!\!\left[ \begin{matrix} \bar{\mathbf{n}}_1^e \boldsymbol{\sigma}(\mathbf{u}, \phi) - \bar{\mathbf{n}}_{21}^e \boldsymbol{\mu}_{,1}(\mathbf{u}, \phi) \\ -\bar{\mathbf{n}}_{22}^e \boldsymbol{\mu}_{,2}(\mathbf{u}, \phi) \end{matrix} \right]\!\!\right] d\Gamma$$
$$= \int_\Omega \mathbf{v}^T \mathbf{b} \, d\Omega + \int_{\partial \Omega} [\![\hat{\boldsymbol{\varepsilon}}^T(\mathbf{v})]\!] \bar{\mathbf{n}}_3 \{\boldsymbol{\mu}(\mathbf{u}, \phi)\} \, d\Gamma$$
$$+ \int_{\partial \Omega} [\![\mathbf{v}^T]\!] \left\{ \begin{matrix} \bar{\mathbf{n}}_1 \boldsymbol{\sigma}(\mathbf{u}, \phi) - \bar{\mathbf{n}}_{21} \boldsymbol{\mu}_{,1}(\mathbf{u}, \phi) \\ -\bar{\mathbf{n}}_{22} \boldsymbol{\mu}_{,2}(\mathbf{u}, \phi) \end{matrix} \right\} d\Gamma \tag{25}$$

$$\int_\Omega \epsilon_0 \mathbf{E}^T(\tau) \mathbf{E}(\phi) \, d\Omega + \int_\Omega \mathbf{E}^T(\tau) \mathbf{P}(\mathbf{u}, \phi) \, d\Omega$$
$$+ \int_{\Gamma_h} [\![\tau]\!] \mathbf{n}^{eT} \{\epsilon_0 \mathbf{E}(\phi) + \mathbf{P}(\mathbf{u}, \phi)\} \, d\Gamma$$
$$+ \int_{\Gamma_h} \{\tau\} \mathbf{n}^{eT} [\![\epsilon_0 \mathbf{E}(\phi) + \mathbf{P}(\mathbf{u}, \phi)]\!] \, d\Gamma$$
$$= \int_\Omega \tau q \, d\Omega - \int_{\partial \Omega} [\![\tau]\!] \mathbf{n}^T \left\{ \begin{matrix} \epsilon_0 \mathbf{E}(\phi) \\ +\mathbf{P}(\mathbf{u}, \phi) \end{matrix} \right\} d\Gamma \tag{26}$$

where

$$\hat{\boldsymbol{\varepsilon}}(\mathbf{u}) = \begin{bmatrix} u_{1,1}, u_{2,2}, u_{1,2}, u_{2,1} \end{bmatrix}^T.$$

Each internal boundary $e \in \Gamma_h$ is shared by two subdomains, i.e., $e = \partial E_1 \cap \partial E_2$. $\mathbf{n}^e$ is defined as the unit vector normal to $e$ and pointing outward from $E_1$, that is, $\mathbf{n}^e = \mathbf{n}|_e^{E_1} = -\mathbf{n}|_e^{E_2}$. The jump operator $[\![ \cdot ]\!]$ and average operator $\{\cdot\}$ are defined as (for $\forall w$):

$$[\![w]\!] = \begin{cases} w|_e^{E_1} - w|_e^{E_2} & e \in \Gamma_h \\ w|_e & e \in \partial \Omega' \end{cases}$$
$$\{w\} = \begin{cases} \frac{1}{2}(w|_e^{E_1} + w|_e^{E_2}) & e \in \Gamma_h \\ w|_e & e \in \partial \Omega \end{cases}.$$

The following boundary term in Eqn. (25) can be integrated by parts:

$$\int_{\partial \Omega} [\![\hat{\boldsymbol{\varepsilon}}^T(\mathbf{v})]\!] \bar{\mathbf{n}}_3 \{\boldsymbol{\mu}(\mathbf{u}, \phi)\} \, d\Gamma$$
$$= \int_{\partial \Omega_d \cup \partial \Omega_R} \hat{\boldsymbol{\varepsilon}}^T(\mathbf{v}) \bar{\mathbf{n}}_4^T \bar{\mathbf{n}}_4 \bar{\mathbf{n}}_3 \boldsymbol{\mu}(\mathbf{u}, \phi) \, d\Gamma$$
$$- \int_{\partial \Omega_u \cup \partial \Omega_Q} \mathbf{v}^T \bar{\mathbf{c}}_1 \bar{\mathbf{s}}_4^T \bar{\mathbf{s}}_4 \bar{\mathbf{n}}_3 \boldsymbol{\mu}_{,1}(\mathbf{u}, \phi) \, d\Gamma$$
$$- \int_{\partial \Omega_u \cup \partial \Omega_Q} \mathbf{v}^T \bar{\mathbf{c}}_2 \bar{\mathbf{s}}_4^T \bar{\mathbf{s}}_4 \bar{\mathbf{n}}_3 \boldsymbol{\mu}_{,2}(\mathbf{u}, \phi) \, d\Gamma. \tag{27}$$

The corresponding matrices in Eqn. (25) – (27) are shown in Appendix B.

When $(\mathbf{u}, \phi)$ are exact solutions, due to continuity conditions, $[\![\boldsymbol{\mu}(\mathbf{u}, \phi)]\!] = \mathbf{0}$, $[\![\bar{\mathbf{n}}_1^e \boldsymbol{\sigma}(\mathbf{u}, \phi) - \bar{\mathbf{n}}_{21}^e \boldsymbol{\mu}_{,1}(\mathbf{u}, \phi) - \bar{\mathbf{n}}_{22}^e \boldsymbol{\mu}_{,2}(\mathbf{u}, \phi)]\!] = \mathbf{0}$ and $[\![\epsilon_0 \mathbf{E}(\phi) + \mathbf{P}(\mathbf{u}, \phi)]\!] = \mathbf{0}$ for $\forall e \in \Gamma_h$. Hence, we can eliminate the corresponding terms in Eqn. (25) and (26). Similarly, $[\![\mathbf{u}]\!] = \mathbf{0}$, $[\![\hat{\boldsymbol{\varepsilon}}(\mathbf{u})]\!] = \mathbf{0}$ and $[\![\phi]\!] = 0$. By adding antithetic terms, the previous equations can be written in a symmetric form.

Furthermore, in order to impose the essential boundary conditions ($\partial \Omega_u$, $\partial \Omega_d$ and $\partial \Omega_\phi$) and the continuity condition across subdomains, we employ the Numerical Flux Correction, which is widely used in Discontinuous Galerkin (DG) Methods [16] to ensure the consistency and stability of the method. Here we use the Interior Penalty (IP) Numerical Flux Corrections. A set of penalty parameters are introduced, in which $\boldsymbol{\eta}_1 = [\eta_{11}, \eta_{12}, \eta_{13}]$ and $\boldsymbol{\eta}_2 = [\eta_{21}, \eta_{22}, \eta_{23}]$ are related to the essential boundary conditions and the interior continuity conditions respectively. The FPM is only stable when the penalty parameters are large enough.

Finally, after substituting the constitutive equations and the following boundary conditions (see Eqn. (4) – (9)):

$$\int_{\partial \Omega_Q} \mathbf{v}^T \begin{pmatrix} \bar{\mathbf{n}}_1 \boldsymbol{\sigma}(\mathbf{u}, \phi) \\ -(\bar{\mathbf{n}}_{21} \bar{\mathbf{c}}_1 \bar{\mathbf{s}}_4^T \bar{\mathbf{s}}_4 \bar{\mathbf{n}}_3) \boldsymbol{\mu}_{,1}(\mathbf{u}, \phi) \\ -(\bar{\mathbf{n}}_{22} - \bar{\mathbf{c}}_2 \bar{\mathbf{s}}_4^T \bar{\mathbf{s}}_4 \bar{\mathbf{n}}_3) \boldsymbol{\mu}_{,2}(\mathbf{u}, \phi) \end{pmatrix} d\Gamma$$
$$= \int_{\partial \Omega_Q} \mathbf{v}^T \widetilde{\mathbf{Q}} \, d\Gamma, \tag{28}$$

$$\int_{\partial\Omega_R} \hat{\boldsymbol{\varepsilon}}^T(\mathbf{v})\bar{\mathbf{n}}_4^T\bar{\mathbf{n}}_4\bar{\mathbf{n}}_3\boldsymbol{\mu}(\mathbf{u},\phi)\mathrm{d}\Gamma$$
$$= \int_{\partial\Omega_R} \hat{\boldsymbol{\varepsilon}}^T(\mathbf{v})\bar{\mathbf{n}}_4^T\widetilde{\mathbf{R}}\mathrm{d}\Gamma, \tag{29}$$

$$\int_{\partial\Omega_\omega} [\![\tau]\!]\mathbf{n}^T\{\epsilon_0\mathbf{E}(\phi) + \mathbf{P}(\mathbf{u},\phi)\}\,\mathrm{d}\Gamma$$
$$= -\int_{\partial\Omega_\omega} \tau\widetilde{\omega}\,\mathrm{d}\Gamma, \tag{30}$$

we can obtain the final symmetric formula for the FPM for flexoelectric problems:

$$\int_\Omega \boldsymbol{\varepsilon}^T(\mathbf{v})\mathbf{D}_{\sigma\varepsilon}\boldsymbol{\varepsilon}(\mathbf{u})\,\mathrm{d}\Omega$$
$$+ \int_\Omega \boldsymbol{\kappa}^T(\mathbf{v})\mathbf{D}_{\mu\kappa}\boldsymbol{\kappa}(\mathbf{u})\,\mathrm{d}\Omega$$
$$- \int_\Omega \left(\boldsymbol{\varepsilon}^T(\mathbf{v})\mathbf{G}_0\boldsymbol{\kappa}(\mathbf{u}) + \boldsymbol{\kappa}^T(\mathbf{v})\mathbf{G}_0^T\boldsymbol{\varepsilon}(\mathbf{u})\right)\mathrm{d}\Omega$$
$$- \int_\Omega \left(\boldsymbol{\varepsilon}^T(\mathbf{v})\mathbf{e} + \boldsymbol{\kappa}^T(\mathbf{v})\mathbf{A}_0^T\right)\mathbf{E}(\phi)\,\mathrm{d}\Omega$$
$$+ \int_{\Gamma_h\cup\partial\Omega_u} \left( [\![\mathbf{v}^T]\!]\begin{pmatrix}\bar{\mathbf{n}}_1{}^e\mathbf{G}_0 \\ -\bar{\mathbf{n}}_{21}{}^e\mathbf{G}_0^T\bar{\mathbf{c}}_3 \\ -\bar{\mathbf{n}}_{22}{}^e\mathbf{G}_0^T\bar{\mathbf{c}}_4\end{pmatrix}\{\boldsymbol{\kappa}(\mathbf{u})\} + \{\boldsymbol{\kappa}^T(\mathbf{v})\}\begin{pmatrix}\mathbf{G}_0^T\bar{\mathbf{n}}_1{}^{eT} \\ -\bar{\mathbf{c}}_3^T\mathbf{G}_0\bar{\mathbf{n}}_{21}{}^{eT} \\ -\bar{\mathbf{c}}_4^T\mathbf{G}_0\bar{\mathbf{n}}_{22}{}^{eT}\end{pmatrix}[\![\mathbf{u}]\!] \right)\mathrm{d}\Gamma$$
$$- \int_{\Gamma_h\cup\partial\Omega_u}\begin{pmatrix}[\![\mathbf{v}^T]\!]\bar{\mathbf{n}}_{21}{}^e\mathbf{A}_0\{\mathbf{E}_{,1}(\phi)\} \\ +[\![\mathbf{v}^T]\!]\bar{\mathbf{n}}_{22}{}^e\mathbf{A}_0\{\mathbf{E}_{,2}(\phi)\}\end{pmatrix}\mathrm{d}\Gamma$$
$$- \int_{\Gamma_h\cup\partial\Omega_u}\begin{pmatrix}[\![\mathbf{v}^T]\!]\bar{\mathbf{n}}_1{}^e\mathbf{D}_{\sigma\varepsilon}\{\boldsymbol{\varepsilon}(\mathbf{u})\} \\ +\{\boldsymbol{\varepsilon}^T(\mathbf{v})\}\mathbf{D}_{\sigma\varepsilon}\bar{\mathbf{n}}_1{}^{eT}[\![\mathbf{u}]\!]\end{pmatrix}\mathrm{d}\Gamma$$
$$+ \int_{\Gamma_h\cup\partial\Omega_u}\begin{pmatrix}[\![\mathbf{v}^T]\!]\bar{\mathbf{n}}_{21}{}^e\mathbf{D}_{\mu\kappa}\{\boldsymbol{\kappa}_{,1}(\mathbf{u})\} \\ +\{\boldsymbol{\kappa}_{,1}{}^T(\mathbf{v})\}\mathbf{D}_{\mu\kappa}\bar{\mathbf{n}}_{21}{}^{eT}[\![\mathbf{u}]\!]\end{pmatrix}\mathrm{d}\Gamma$$
$$+ \int_{\Gamma_h\cup\partial\Omega_u}\begin{pmatrix}[\![\mathbf{v}^T]\!]\bar{\mathbf{n}}_{22}{}^e\mathbf{D}_{\mu\kappa}\{\boldsymbol{\kappa}_{,2}(\mathbf{u})\} \\ +\{\boldsymbol{\kappa}_{,2}{}^T(\mathbf{v})\}\mathbf{D}_{\mu\kappa}\bar{\mathbf{n}}_{22}{}^{eT}[\![\mathbf{u}]\!]\end{pmatrix}\mathrm{d}\Gamma$$
$$+ \int_{\Gamma_h\cup\partial\Omega_u} [\![\mathbf{v}^T]\!]\bar{\mathbf{n}}_1{}^e\mathbf{e}\{\mathbf{E}(\phi)\}\mathrm{d}\Gamma$$
$$- \int_{\Gamma_h}\begin{pmatrix}[\![\hat{\boldsymbol{\varepsilon}}^T(\mathbf{v})]\!]\bar{\mathbf{n}}_3{}^e\mathbf{D}_{\mu\kappa}\{\boldsymbol{\kappa}(\mathbf{u})\} \\ +\{\boldsymbol{\kappa}^T(\mathbf{v})\}\mathbf{D}_{\mu\kappa}\bar{\mathbf{n}}_3{}^{eT}[\![\hat{\boldsymbol{\varepsilon}}(\mathbf{u})]\!]\end{pmatrix}\mathrm{d}\Gamma$$
$$+ \int_{\Gamma_h}\begin{pmatrix}[\![\hat{\boldsymbol{\varepsilon}}^T(\mathbf{v})]\!]\bar{\mathbf{n}}_3{}^e\mathbf{G}_0^T\{\boldsymbol{\varepsilon}(\mathbf{u})\} \\ +\{\boldsymbol{\varepsilon}^T(\mathbf{v})\}\mathbf{G}_0\bar{\mathbf{n}}_3{}^{eT}[\![\hat{\boldsymbol{\varepsilon}}(\mathbf{u})]\!]\end{pmatrix}\mathrm{d}\Gamma$$
$$+ \int_{\Gamma_h} [\![\hat{\boldsymbol{\varepsilon}}^T(\mathbf{v})]\!]\bar{\mathbf{n}}_3{}^e\mathbf{A}_0\{\mathbf{E}(\phi)\}\mathrm{d}\Gamma$$
$$+ \int_{\partial\Omega_u}\begin{pmatrix}\mathbf{v}^T\bar{\mathbf{c}}_1\bar{\mathbf{s}}_4^T\bar{\mathbf{s}}_4\bar{\mathbf{n}}_3\mathbf{D}_{\mu\kappa}\boldsymbol{\kappa}_{,1}(\mathbf{u}) \\ +\boldsymbol{\kappa}_{,1}{}^T(\mathbf{v})\mathbf{D}_{\mu\kappa}\bar{\mathbf{n}}_3{}^{eT}\bar{\mathbf{s}}_4^T\bar{\mathbf{s}}_4\bar{\mathbf{c}}_1^T\mathbf{u}\end{pmatrix}\mathrm{d}\Gamma$$
$$+ \int_{\partial\Omega_u}\begin{pmatrix}\mathbf{v}^T\bar{\mathbf{c}}_2\bar{\mathbf{s}}_4^T\bar{\mathbf{s}}_4\bar{\mathbf{n}}_3\mathbf{D}_{\mu\kappa}\boldsymbol{\kappa}_{,2}(\mathbf{u}) \\ +\boldsymbol{\kappa}_{,2}{}^T(\mathbf{v})\mathbf{D}_{\mu\kappa}\bar{\mathbf{n}}_3^T\bar{\mathbf{s}}_4^T\bar{\mathbf{s}}_4\bar{\mathbf{c}}_2^T\mathbf{u}\end{pmatrix}\mathrm{d}\Gamma$$
$$- \int_{\partial\Omega_u}\left(\mathbf{v}^T\begin{pmatrix}\bar{\mathbf{c}}_1\bar{\mathbf{s}}_4^T\bar{\mathbf{s}}_4\bar{\mathbf{n}}_3\mathbf{G}_0^T\bar{\mathbf{c}}_3 \\ +\bar{\mathbf{c}}_2\bar{\mathbf{s}}_4^T\bar{\mathbf{s}}_4\bar{\mathbf{n}}_3\mathbf{G}_0^T\bar{\mathbf{c}}_4\end{pmatrix}\boldsymbol{\kappa}(\mathbf{u}) + \boldsymbol{\kappa}^T(\mathbf{v})\begin{pmatrix}\bar{\mathbf{c}}_3^T\mathbf{G}_0\bar{\mathbf{n}}_3^T\bar{\mathbf{s}}_4^T\bar{\mathbf{s}}_4\bar{\mathbf{c}}_1^T \\ +\bar{\mathbf{c}}_4^T\mathbf{G}_0\bar{\mathbf{n}}_3^T\bar{\mathbf{s}}_4^T\bar{\mathbf{s}}_4\bar{\mathbf{c}}_2^T\end{pmatrix}\mathbf{u}\right)\mathrm{d}\Gamma$$
$$- \int_{\partial\Omega_u}\begin{pmatrix}\mathbf{v}^T\bar{\mathbf{c}}_1\bar{\mathbf{s}}_4^T\bar{\mathbf{s}}_4\bar{\mathbf{n}}_3\mathbf{A}_0\mathbf{E}_{,1}(\phi) \\ +\mathbf{v}^T\bar{\mathbf{c}}_2\bar{\mathbf{s}}_4^T\bar{\mathbf{s}}_4\bar{\mathbf{n}}_3\mathbf{A}_0\mathbf{E}_{,2}(\phi)\end{pmatrix}\mathrm{d}\Gamma$$
$$- \int_{\partial\Omega_d}\begin{pmatrix}\hat{\boldsymbol{\varepsilon}}^T(\mathbf{v})\bar{\mathbf{n}}_4^T\bar{\mathbf{n}}_4\bar{\mathbf{n}}_3\mathbf{D}_{\mu\kappa}\boldsymbol{\kappa}(\mathbf{u}) \\ +\boldsymbol{\kappa}^T(\mathbf{v})\mathbf{D}_{\mu\kappa}\bar{\mathbf{n}}_3^T\bar{\mathbf{n}}_4^T\bar{\mathbf{n}}_4\hat{\boldsymbol{\varepsilon}}(\mathbf{u})\end{pmatrix}\mathrm{d}\Gamma$$
$$+ \int_{\partial\Omega_d}\begin{pmatrix}\hat{\boldsymbol{\varepsilon}}^T(\mathbf{v})\bar{\mathbf{n}}_4^T\bar{\mathbf{n}}_4\bar{\mathbf{n}}_3\mathbf{G}_0^T\boldsymbol{\varepsilon}(\mathbf{u}) \\ +\boldsymbol{\varepsilon}^T(\mathbf{v})\mathbf{G}_0\bar{\mathbf{n}}_3^T\bar{\mathbf{n}}_4^T\bar{\mathbf{n}}_4\hat{\boldsymbol{\varepsilon}}(\mathbf{u})\end{pmatrix}\mathrm{d}\Gamma$$
$$+ \int_{\partial\Omega_d} \hat{\boldsymbol{\varepsilon}}^T(\mathbf{v})\bar{\mathbf{n}}_4^T\bar{\mathbf{n}}_4\bar{\mathbf{n}}_3\mathbf{A}_0\mathbf{E}(\phi)\,\mathrm{d}\Gamma$$
$$- \int_{\Gamma_h\cup\partial\Omega_\phi}\begin{pmatrix}\{\boldsymbol{\varepsilon}^T(\mathbf{v})\}\mathbf{e} \\ +\{\boldsymbol{\kappa}^T(\mathbf{v})\}\mathbf{A}_0\end{pmatrix}\mathbf{n}^e[\![\phi]\!]\mathrm{d}\Gamma$$
$$+ \int_{\partial\Omega_u}\frac{\eta_{11}}{h_e}\mathbf{v}^T\mathbf{u}\mathrm{d}\Gamma + \int_{\Gamma_h}\frac{\eta_{21}}{h_e}[\![\mathbf{v}^T]\!][\![\mathbf{u}]\!]\mathrm{d}\Gamma$$
$$+ \int_{\partial\Omega_d} \eta_{12}\,h_e\hat{\boldsymbol{\varepsilon}}^T(\mathbf{v})\bar{\mathbf{n}}_4^T\bar{\mathbf{n}}_4\hat{\boldsymbol{\varepsilon}}(\mathbf{u})\mathrm{d}\Gamma$$
$$+ \int_{\Gamma_h} \eta_{22}\,h_e[\![\hat{\boldsymbol{\varepsilon}}^T(\mathbf{v})]\!]\bar{\mathbf{n}}_4^T\bar{\mathbf{n}}_4[\![\hat{\boldsymbol{\varepsilon}}(\mathbf{u})]\!]\mathrm{d}\Gamma$$
$$= \int_\Omega \mathbf{v}^T\mathbf{b}\mathrm{d}\Omega + \int_{\partial\Omega_Q}\mathbf{v}^T\widetilde{\mathbf{Q}}\,\mathrm{d}\Gamma$$
$$+ \int_{\partial\Omega_R} \hat{\boldsymbol{\varepsilon}}^T(\mathbf{v})\bar{\mathbf{n}}_4^T\widetilde{\mathbf{R}}\mathrm{d}\Gamma - \int_{\partial\Omega_u}\boldsymbol{\varepsilon}^T(\mathbf{v})\mathbf{D}_{\sigma\varepsilon}\bar{\mathbf{n}}_1{}^{eT}\widetilde{\mathbf{u}}\mathrm{d}\Gamma$$
$$+ \int_{\partial\Omega_u}\boldsymbol{\kappa}^T(\mathbf{v})\begin{pmatrix}\mathbf{G}_0^T\bar{\mathbf{n}}_1{}^{eT} - \bar{\mathbf{c}}_3^T\mathbf{G}_0\bar{\mathbf{n}}_{21}{}^{eT} \\ -\bar{\mathbf{c}}_4^T\mathbf{G}_0\bar{\mathbf{n}}_{22}{}^{eT}\end{pmatrix}\widetilde{\mathbf{u}}\mathrm{d}\Gamma$$
$$+ \int_{\partial\Omega_u}\left[\begin{matrix}\boldsymbol{\kappa}_{,1}{}^T(\mathbf{v})\mathbf{D}_{\mu\kappa}(\bar{\mathbf{n}}_{21}{}^{eT} + \bar{\mathbf{n}}_3{}^T\bar{\mathbf{s}}_4^T\bar{\mathbf{s}}_4\bar{\mathbf{c}}_1^T) \\ +\boldsymbol{\kappa}_{,2}{}^T(\mathbf{v})\mathbf{D}_{\mu\kappa}(\bar{\mathbf{n}}_{22}{}^{eT} + \bar{\mathbf{n}}_3^T\bar{\mathbf{s}}_4^T\bar{\mathbf{s}}_4\bar{\mathbf{c}}_2^T)\end{matrix}\right]\widetilde{\mathbf{u}}\mathrm{d}\Gamma$$
$$- \int_{\partial\Omega_d}\begin{pmatrix}\boldsymbol{\kappa}^T(\mathbf{v})\mathbf{D}_{\mu\kappa} \\ -\boldsymbol{\varepsilon}^T(\mathbf{v})\mathbf{G}_0\end{pmatrix}\bar{\mathbf{n}}_3^T\bar{\mathbf{n}}_4^T\check{\mathbf{d}}d\Gamma$$

$$-\int_{\partial\Omega_\phi}\begin{pmatrix}\boldsymbol{\varepsilon}^T(\mathbf{v})\mathbf{e}\\+\boldsymbol{\kappa}^T(\mathbf{v})\mathbf{A}_0\end{pmatrix}\mathbf{n}\tilde{\phi}\,\mathrm{d}\Gamma+\int_{\partial\Omega_u}\frac{\eta_{11}}{h_e}\mathbf{v}^T\tilde{\mathbf{u}}\,\mathrm{d}\Gamma$$
$$+\int_{\partial\Omega_d}\eta_{12}\,h_e\hat{\boldsymbol{\varepsilon}}^T(\mathbf{v})\bar{\mathbf{n}}_4^{\ T}\tilde{\mathbf{d}}\,\mathrm{d}\Gamma, \tag{31}$$

$$-\int_\Omega \mathbf{E}^T(\tau)\bar{\boldsymbol{\kappa}}\mathbf{E}(\phi)\,\mathrm{d}\Omega$$
$$-\int_\Omega \mathbf{E}^T(\tau)\left(\mathbf{e}^T\boldsymbol{\varepsilon}(\mathbf{u})+\mathbf{A}_0^{\ T}\boldsymbol{\kappa}(\mathbf{u})\right)\mathrm{d}\Omega$$
$$+\int_{\Gamma_h\cup\partial\Omega_u}\{\mathbf{E}^T(\tau)\}\mathbf{e}^T\bar{\mathbf{n}}_1^{\ eT}[\![\mathbf{u}]\!]\mathrm{d}\Gamma$$
$$-\int_{\Gamma_h\cup\partial\Omega_u}\begin{pmatrix}\{\mathbf{E}_{,1}^{\ T}(\tau)\}\mathbf{A}_0^{\ T}\bar{\mathbf{n}}_{21}^{\ eT}[\![\mathbf{u}]\!]\\+\{\mathbf{E}_{,2}^{\ T}(\tau)\}\mathbf{A}_0^{\ T}\bar{\mathbf{n}}_{22}^{\ eT}[\![\mathbf{u}]\!]\end{pmatrix}\mathrm{d}\Gamma$$
$$-\int_{\Gamma_h\cup\partial\Omega_\phi}\begin{pmatrix}[\![\tau]\!]\mathbf{n}^{eT}\bar{\boldsymbol{\kappa}}\{\mathbf{E}(\phi)\}\\+\{\mathbf{E}^T(\tau)\}\bar{\boldsymbol{\kappa}}\mathbf{n}^e[\![\phi]\!]\end{pmatrix}\mathrm{d}\Gamma$$
$$-\int_{\Gamma_h\cup\partial\Omega_\phi}[\![\tau]\!]\mathbf{n}^{eT}\begin{pmatrix}\mathbf{e}^T\{\boldsymbol{\varepsilon}(\mathbf{u})\}\\+\mathbf{A}_0^{\ T}\{\boldsymbol{\kappa}(\mathbf{u})\}\end{pmatrix}\mathrm{d}\Gamma$$
$$+\int_{\Gamma_h}\{\mathbf{E}^T(\tau)\}\mathbf{A}_0^{\ T}\bar{\mathbf{n}}_3^{\ eT}[\![\hat{\boldsymbol{\varepsilon}}(\mathbf{u})]\!]\,\mathrm{d}\Gamma$$
$$-\int_{\partial\Omega_u}\begin{pmatrix}\mathbf{E}_{,1}^{\ T}(\tau)\mathbf{A}_0^{\ T}\bar{\mathbf{n}}_3^{\ T}\bar{\mathbf{s}}_4^{\ T}\bar{\mathbf{s}}_4\bar{\mathbf{c}}_1^{\ T}\\+\mathbf{E}_{,2}^{\ T}(\tau)\mathbf{A}_0^{\ T}\bar{\mathbf{n}}_3^{\ T}\bar{\mathbf{s}}_4^{\ T}\bar{\mathbf{s}}_4\bar{\mathbf{c}}_2^{\ T}\end{pmatrix}\mathbf{u}\,\mathrm{d}\Gamma$$
$$+\int_{\partial\Omega_d}\mathbf{E}^T(\tau)\mathbf{A}_0^{\ T}\bar{\mathbf{n}}_3^{\ T}\bar{\mathbf{n}}_4^{\ T}\hat{\boldsymbol{\varepsilon}}(\mathbf{u})\,\mathrm{d}\Gamma$$
$$+\int_{\partial\Omega_\phi}\frac{\eta_{13}}{h_e}\tau\phi\,\mathrm{d}\Gamma+\int_{\Gamma_h}\frac{\eta_{23}}{h_e}[\![\tau]\!][\![\phi]\!]\mathrm{d}\Gamma$$
$$=-\int_\Omega \tau q\,\mathrm{d}\Omega-\int_{\partial\Omega_\omega}\tau\tilde{\omega}\,\mathrm{d}\Gamma+\int_{\partial\Omega_u}\mathbf{E}^T(\tau)\mathbf{e}^T\bar{\mathbf{n}}_1^{\ T}\tilde{\mathbf{u}}\,\mathrm{d}\Gamma$$
$$-\int_{\partial\Omega_u}\begin{bmatrix}\mathbf{E}_{,1}^{\ T}(\tau)\mathbf{A}_0^{\ T}\left(\bar{\mathbf{n}}_{21}^{\ T}+\bar{\mathbf{n}}_3^{\ T}\bar{\mathbf{s}}_4^{\ T}\bar{\mathbf{s}}_4\bar{\mathbf{c}}_1^{\ T}\right)\\+\mathbf{E}_{,2}^{\ T}(\tau)\mathbf{A}_0^{\ T}\left(\bar{\mathbf{n}}_{22}^{\ T}+\bar{\mathbf{n}}_3^{\ T}\bar{\mathbf{s}}_4^{\ T}\bar{\mathbf{s}}_4\bar{\mathbf{c}}_2^{\ T}\right)\end{bmatrix}\tilde{\mathbf{u}}\,\mathrm{d}\Gamma$$
$$+\int_{\partial\Omega_d}\mathbf{E}^T(\tau)\mathbf{A}_0^{\ T}\bar{\mathbf{n}}_3^{\ T}\bar{\mathbf{n}}_4^{\ T}\tilde{\mathbf{d}}\,\mathrm{d}\Gamma-\int_{\partial\Omega_\phi}\mathbf{E}^T(\tau)\bar{\boldsymbol{\kappa}}\mathbf{n}\tilde{\phi}\,\mathrm{d}\Gamma$$
$$+\int_{\partial\Omega_\phi}\frac{\eta_{13}}{h_e}\tau\tilde{\phi}\,\mathrm{d}\Gamma. \tag{32}$$

where the normal and constant matrices $\bar{\mathbf{n}}_i$, $\bar{\mathbf{c}}_i$ are shown in Appendix B. $h_e$ is a boundary-dependent parameter with the unit of length. Here it is defined as the distance between the points in subdomains sharing the boundary for $e \in \Gamma_h$, and the smallest distance between the internal point and the external boundary for $e \in \partial\Omega$. The penalty parameters are independent of the boundary size, in which $(\eta_{11},\eta_{12},\eta_{21},\eta_{22})$ have the same unit of the Young's modulus, and $(\eta_{13},\eta_{23})$ have the same unit of the permittivity of the material.

As has been stated, the method is only stable when the penalty parameter $\boldsymbol{\eta}_2$ is large enough. However, unlike the DG methods, in which the shape functions are completely independent, a small penalty parameter $\boldsymbol{\eta}_2$ is enough to stabilize the present FPM. On the other hand, the accuracy may decrease if $\boldsymbol{\eta}_2$ is too large. There is no upper limit for $\boldsymbol{\eta}_1$ which corresponds to the essential boundary conditions.

At last, the formula of the FPM can be converted into the following form:

$$\begin{bmatrix}\mathbf{K}_{uu} & \mathbf{K}_{u\phi}\\ \mathbf{K}_{u\phi}^T & \mathbf{K}_{\phi\phi}\end{bmatrix}\begin{bmatrix}\bar{\mathbf{u}}\\ \bar{\boldsymbol{\phi}}\end{bmatrix}=\begin{bmatrix}\mathbf{f}_u\\ \mathbf{f}_\phi\end{bmatrix},\quad \text{or } \mathbf{K}\bar{\mathbf{x}}=\mathbf{f}, \tag{33}$$

where $(\bar{\mathbf{u}},\bar{\boldsymbol{\phi}})$ are the global nodal displacement and electric potential vectors. $\mathbf{K}$ is the global stiffness matrix, assembled by a series of point stiffness matrix ($\mathbf{K}_E$) and boundary stiffness matrices ($\mathbf{K}_h$, $\mathbf{K}_u$, $\mathbf{K}_Q$, $\mathbf{K}_d$, $\mathbf{K}_R$, $\mathbf{K}_\phi$ and $\mathbf{K}_\omega$ corresponding to $\Gamma_h$, $\partial\Omega_u$, $\partial\Omega_Q$, $\partial\Omega_d$, $\partial\Omega_R$, $\partial\Omega_\phi$ and $\partial\Omega_\omega$ respectively). $\mathbf{f}$ is the global load vector. When establishing these stiffness matrices and load vectors, we use Gaussian quadrature rule. $2\times 2$ integration points are used in each subdomain, while only one integration point is used on each internal or external boundary. The numerical implementation is omitted here. The final stiffness matrix is sparse and symmetric, which is good for modeling complex problems with a large number of DOFs.

## 5. NUMERICAL EXAMPLES

### 5.1 Hollow cylinder

The first example is an infinite length flexoelectric tube. As shown in Fig. 3, this is a plane strain problem with axisymmetric boundary conditions.

The material properties are: Young's modulus $E = 139\text{GPa}$, Poisson's ratio $\nu = 0.3$, internal material length $l = 2\mu m$, flexoelectric coefficients $\bar{\mu}_{11} = \bar{\mu}_{12} = \bar{\mu}_{44} = 1\times 10^{-6}\,\text{C/m}$ and permittivity of the dielectric $\bar{\kappa}_{11} = \bar{\kappa}_{33} = 1\times 10^{-9}\,\text{F/m}$ (see Appendix A). The material is non-piezoelectric. The geometric parameters are: $r_i = 10\mu m$, $r_o = 20\mu m$. And the boundary conditions are given as: $u_i = 0.045\mu m$, $u_o = 0.05\mu m$, $\phi_i = 0V$, $\phi_o = 1V$.

In the FPM, according to the symmetry, only a quarter of the entire domain is considered. Symmetric boundary conditions are applied. 1260 uniform points are employed. The computational parameters are given as: $c_0 = \sqrt{10}$, $\eta_{11} = 1\times 10^{10}E$, $\eta_{13} = 1\times 10^{10}\bar{\kappa}_{33}$, $\eta_{21} = 2.0E$, $\eta_{22} = 100E$, $\eta_{23} = 0$.

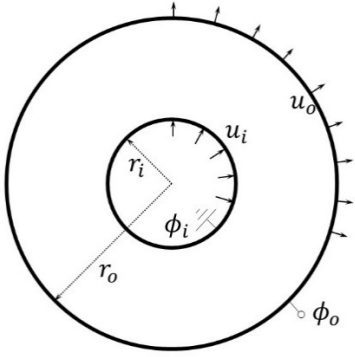

**FIGURE 3:** AN INFINITE LENGTH TUBE WITH AN AXISYMMETRIC CROSS SECTION.

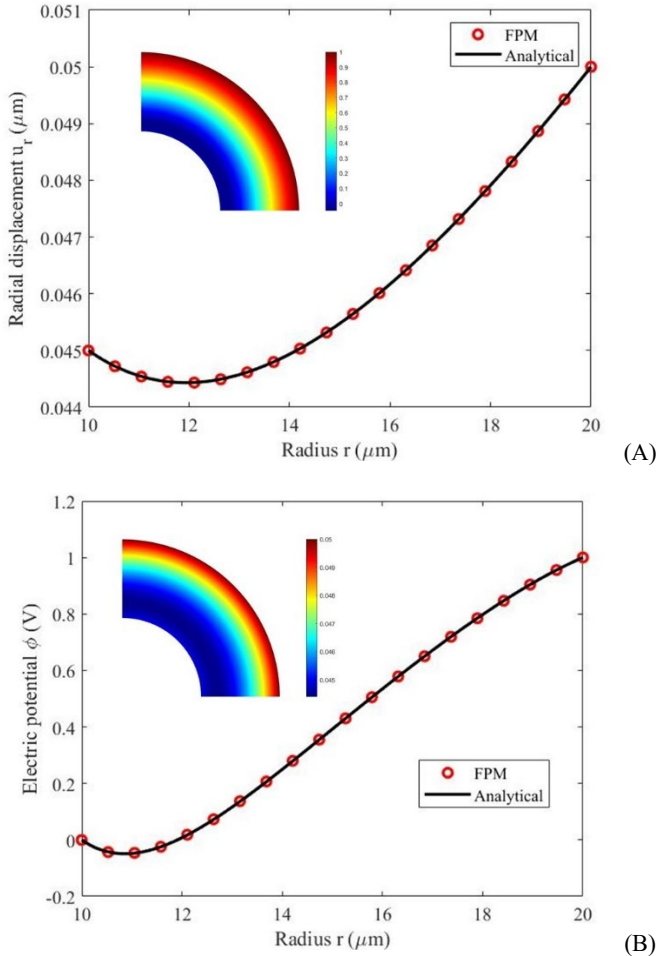

**FIGURE 4:** THE COMPUTED SOLUTION. (A) DISTRIBUTION OF RADIAL DISPLACEMENT. (B) DISTRIBUTION OF ELECTRIC POTENTIAL.

The same example is also considered in [2] and [4] using the mixed FEM. The computed solution by the FPM is presented in Fig. 4, comparing with the analytical solution given by Mao and Purohit [17]. As can be seen, the numerical solution shows excellent agreement with the analytical result. We define the relative errors of displacement and electric potential as:

$$e_u = \frac{\|\mathbf{u}^h - \mathbf{u}\|_{L^2}}{\|\mathbf{u}\|_{L^2}}, \quad e_\phi = \frac{\|\phi^h - \phi\|_{L^2}}{\|\phi\|_{L^2}}, \quad (34)$$

where

$$\|\mathbf{u}\|_{L^2} = \int_\Omega \mathbf{u}^T \mathbf{u} \, d\Omega, \quad \|\phi\|_{L^2} = \int_\Omega \phi^2 \, d\Omega$$

In this example, the relative errors are $e_u = 3.2 \times 10^{-5}$ and $e_\phi = 8.4 \times 10^{-4}$. Apparently, the present FPM presents significantly high accuracy in this benchmark example.

**5.2 2D block subject to a concentrated load**

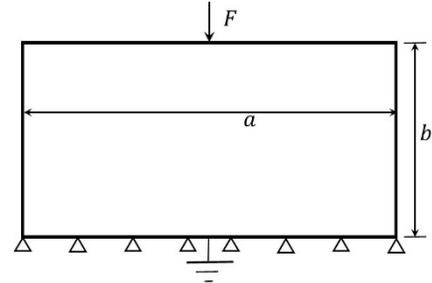

**FIGURE 5:** 2D BLOCK SUBJECT TO A POINT LOAD.

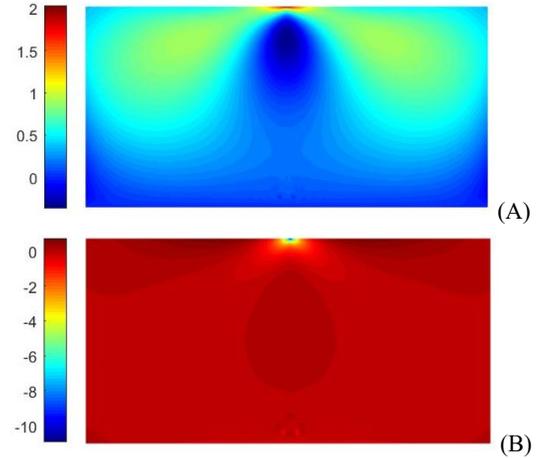

**FIGURE 6:** THE COMPUTED SOLUTION. (A) DISTRIBUTION OF ELECTRIC POTENTIAL $\phi$ (V). (B) DISTRIBUTION OF ELECTRIC FIELD $E_2$ (V/$\mu m$).

Second, we consider a 2D plane strain model of a block subject to a concentrated load. As shown in Fig. 5, a significant strain gradient and electric field can be expected. The example is initiated from the flexoelectric effect observed in some atomic force microscope experiments [2].

The material properties are the same as the first example. The geometric parameters are: $a = 20\mu m$, $b = 10\mu m$. The concentrated force $F = 100\mu N$, applied uniformly in a $200nm$ width area to avoid singularity. 3200 points are utilized in the FPM. Both Voronoi Diagram partition and quadrilateral partition (generated by the preprocessing module of ABAQUS) are employed. The result shows great consistency between the different partitions. For simplicity, only the solution of the quadrilateral partition is presented in the paper.

The computational parameters in the FPM are: $c_0 = \sqrt{20}$, $\eta_{11} = 1 \times 10^{10} E$, $\eta_{13} = 1 \times 10^{10} \bar{\kappa}_{33}$, $\eta_{21} = 1.0E$, $\eta_{22} = 50E$, $\eta_{23} = 0$. Figure 6 exhibits the numerical solution of the electric potential $\phi$ and the vertical electric field $E_2$. The result shows great consistency with [2] in which the example is studied by the mixed FEM.

### 5.3 Simply supported truncated pyramid

The third example is a truncated pyramid subject to a uniform force (see Fig. 7). In this example, the material is anisotropic. And the piezoelectric effect is taken into consideration. The material properties are: $E = 100 \text{GPa}$, $\nu = 0.37$, $\bar{\kappa}_{11} = 11 \times 10^{-9}$ F/m, $\bar{\kappa}_{33} = 12.48 \times 10^{-9}$ F/m, $\bar{\mu}_{12} = 1 \times 10^{-6}$ C/m, $\bar{\mu}_{11} = \bar{\mu}_{44} = 0$ and piezoelectric coefficients $e_{33} = -4.4$ C/m$^2$, $e_{15} = e_{31} = 0$. The strain gradient effect is neglected ($l = 0$). The geometric parameters are: $a_1 = 750 \mu m$, $a_2 = 2250 \mu m$, $b = 750 \mu m$. The force $F = 450 kN$ is distributed uniformly on the top and the bottom surface. The electric potential on the top surface is fixed to zero while on the bottom surface, it is subject to a constant but priori unknown value $V$.

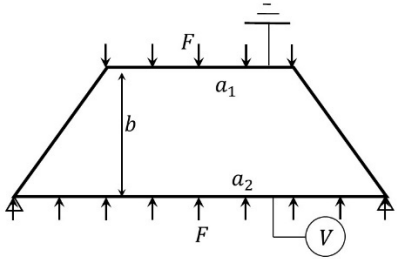

**FIGURE 7:** SIMPLY SUPPORTED TRUNCATED PYRAMID.

With 306 uniform points and quadrilateral partition used, the numerical solutions of the electric potential $\phi$ and mechanical strain $\varepsilon_{22}$ are presented in Fig. 8. The computational parameters $c_0 = \sqrt{20}$, $\eta_{11} = 1 \times 10^{10} E$, $\eta_{13} = 1 \times 10^{10} \bar{\kappa}_{33}$, $\eta_{21} = 1.0E$, $\eta_{22} = 0$, $\eta_{23} = \bar{\kappa}_{33}$. The deformation of the truncated pyramid shows a bending component as expected. The mechanical strain gradient is significant, as well as the nonhomogeneous distribution of the electric potential. The same example is also studied in [2, 10] based on LME meshfree approximation, in which the numerical result agrees well with the current result achieved by the FPM.

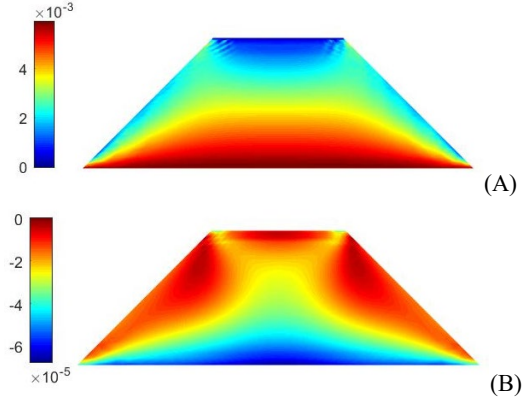

**FIGURE 8:** THE COMPUTED SOLUTION. (A) DISTRIBUTION OF ELECTRIC POTENTIAL $\phi$ ($V$). (B) DISTRIBUTION OF STRAIN $\varepsilon_{22}$.

### 6. CONCLUSION

A new computational approach is developed for analyzing piezoelectric and flexoelectric effects in 2D solids. The purely meshless Fragile Points Method (FPM) based on Galerkin weak-form formulation is superior to the previous meshless methods (e.g., EFG, MLPG, etc.) as it has a much simpler numerical integration scheme based on a polynomial shape function, and results in symmetric and sparse matrices. Numerical examples are carried out for validation. In conclusion, the present FPM is accurate, efficient, and has a great potential in analyzing flexoelectric problems in complex geometries.

### REFERENCES

[1] Wang, Bo, Gu, Yijia, Zhang, Shujun, and Chen, Long-Qing. "Flexoelectricity in solids: Progress, challenges, and perspectives." *Progress in Materials Science* Vol. 106 (2019): pp. 100570.

[2] Zhuang, Xiaoying, Nguyen, Binh Huy, Nanthakumar, Subbiah Srivilliputtur, Tran, Thai Quoc, Alajlan, Naif, and Rabczuk, Timon. "Computational Modeling of Flexoelectricity—A Review." *Energies* Vol. 13 No. 6 (2020): pp. 1326.

[3] Mao, Sheng, Purohit, Prashant K., and Aravas, Nikolaos. "Mixed finite-element formulations in piezoelectricity and flexoelectricity." *Proceedings of the Royal Society A: Mathematical, Physical and Engineering Sciences* Vol. 472 (2016): pp. 20150879.

[4] Deng, Feng, Deng, Qian, Yu, Wenshan, and Shen, Shengping. "Mixed finite elements for flexoelectric solids." *Journal of Applied Mechanics* Vol. 84 No. 8 (2017): pp. 081004.

[5] Maranganti, R., Sharma, N. D., and Sharma, P.. "Electromechanical coupling in nonpiezoelectric materials due to nanoscale nonlocal size effects: Green's function solutions and embedded inclusions." *Physical Review B* Vol. 74 No. 1 (2006): pp. 014110.

[6] Hu, ShuLing, and Shen, ShengPing. "Variational principles and governing equations in nano-dielectrics with the

flexoelectric effect." *Science China Physics, Mechanics and Astronomy* Vol. 53 No. 8 (2010): pp. 1497-1504.

[7] Deng, Qian, Liu, Liping, and Sharma, Pradeep. "Flexoelectricity in soft materials and biological membranes." *Journal of the Mechanics and Physics of Solids* Vol. 62 (2014): pp. 209-227.

[8] Yvonnet, Julien, and Liu, L. P.. "A numerical framework for modeling flexoelectricity and Maxwell stress in soft dielectrics at finite strains." *Computer Methods in Applied Mechanics and Engineering* Vol. 313 (2017): pp. 450-482.

[9] Sladek, Jan, Sladek, Vladimir, Wünsche, Michael, and Zhang, Chuanzeng. "Effects of electric field and strain gradients on cracks in piezoelectric solids." *European Journal of Mechanics-A/Solids* Vol. 71 (2018): pp.187-198.

[10] Abdollahi, Amir, Peco, Christian, Millan, Daniel, Arroyo, Marino, and Arias, Irene. "Computational evaluation of the flexoelectric effect in dielectric solids." *Journal of Applied Physics* Vol. 116 No. 9 (2014): pp. 093502.

[11] He, Bo, Javvaji, Brahmanandam, and Zhuang, Xiaoying. "Characterizing Flexoelectricity in Composite Material Using the Element-Free Galerkin Method." *Energies* Vol. 12 No. 2 (2019): pp. 271.

[12] Tang, Z., Shen, S., and Atluri, S. N.. "Analysis of materials with strain-gradient effects: A meshless local Petrov-Galerkin (MLPG) approach, with nodal displacements only." *Computer Modeling in Engineering and Sciences* Vol. 4 No. 1 (2003): pp. 177-196.

[13] Hillman, Michael, and Chen, Jiun-Shyan. "An accelerated, convergent, and stable nodal integration in Galerkin meshfree methods for linear and nonlinear mechanics." *International Journal for Numerical Methods in Engineering* Vol. 107 No. 7 (2016): pp. 603-630.

[14] Duan, Qinglin, Li, Xikui, Zhang, Hongwu, and Belytschko, Ted. "Second-order accurate derivatives and integration schemes for meshfree methods." *International Journal for Numerical Methods in Engineering* Vol. 92 No. 4 (2012): pp. 399-424.

[16] Shu, C., Ding, H., and Yeo, K. S.. "Local radial basis function-based differential quadrature method and its application to solve two-dimensional incompressible Navier–Stokes equations." *Computer methods in applied mechanics and engineering* Vol. 192 No. 7-8 (2003): pp. 941-954.

[17] Mozolevski, Igor, Süli, Endre, and Bösing, Paulo R. "hp-version a priori error analysis of interior penalty discontinuous Galerkin finite element approximations to the biharmonic equation." *Journal of Scientific Computing* Vol. 30 No. 3 (2007): pp. 465-491.

[16] Mao, Sheng, and Purohit, Prashant K.. "Insights into flexoelectric solids from strain-gradient elasticity." *Journal of Applied Mechanics* Vol. 81 No. 8 (2014): pp. 081004.

**Appendix A: Material Properties**

The material property matrices are given as:

$$\bar{D}_{\sigma\varepsilon} = \begin{bmatrix} \lambda + 2G & \lambda & 0 \\ \lambda & \lambda + 2G & 0 \\ 0 & 0 & G \end{bmatrix},$$

$$\bar{D}_{\mu\kappa} = l^2 \cdot$$

$$\begin{bmatrix} \lambda + 2G & 0 & 0 & \frac{\lambda}{2} & 0 & 0 \\ 0 & \lambda + 2G & \frac{\lambda}{2} & 0 & 0 & 0 \\ 0 & \frac{\lambda}{2} & \frac{\lambda+3G}{4} & 0 & 0 & \frac{G}{2} \\ \frac{\lambda}{2} & 0 & 0 & \frac{\lambda+3G}{4} & \frac{G}{2} & 0 \\ 0 & 0 & 0 & \frac{G}{2} & G & 0 \\ 0 & 0 & \frac{G}{2} & 0 & 0 & G \end{bmatrix},$$

$$\bar{\kappa} = \begin{bmatrix} \bar{\kappa}_{11} & 0 \\ 0 & \bar{\kappa}_{33} \end{bmatrix}, \quad \mathbf{e} = \begin{bmatrix} 0 & 0 & e_{15} \\ e_{31} & e_{33} & 0 \end{bmatrix}, \quad (35)$$

$$\mathbf{A}_0 = \begin{bmatrix} \bar{\mu}_{11} & 0 & 0 & \frac{\bar{\mu}_{12}+\bar{\mu}_{44}}{2} & \bar{\mu}_{44} & 0 \\ 0 & \bar{\mu}_{11} & \frac{\bar{\mu}_{12}+\bar{\mu}_{44}}{2} & 0 & 0 & \bar{\mu}_{44} \end{bmatrix}^{\mathrm{T}},$$

where $(\lambda, G)$ are Lamé parameters, $l$ is the internal material length, $\bar{\kappa}$, $\mathbf{e}$, and $\mathbf{A}_0$ are related to the permittivity of the dielectric, the piezoelectric tensor and the flexoelectric tensor respectively.

**Appendix B: Matrices in Numerical Implementation**

$$\bar{\mathbf{c}}_1 = \begin{bmatrix} 1 & 0 & 0 & 0 \\ 0 & 0 & 0 & 1 \end{bmatrix}, \bar{\mathbf{c}}_2 = \begin{bmatrix} 0 & 0 & 1 & 0 \\ 0 & 1 & 0 & 0 \end{bmatrix},$$

$$\bar{\mathbf{c}}_3 = \begin{bmatrix} 1 & 0 & 0 & 0 & 0 & 0 \\ 0 & 0 & 0 & 1/2 & 0 & 0 \\ 0 & 0 & 1/2 & 0 & 0 & 1 \end{bmatrix},$$

$$\bar{\mathbf{c}}_4 = \begin{bmatrix} 0 & 0 & 1/2 & 0 & 0 & 0 \\ 0 & 1 & 0 & 0 & 0 & 0 \\ 0 & 0 & 0 & 1/2 & 1 & 0 \end{bmatrix},$$

$$\bar{\mathbf{n}}_1 = \begin{bmatrix} n_1 & 0 & n_2 \\ 0 & n_2 & n_1 \end{bmatrix}, \quad (36)$$

$$\bar{\mathbf{n}}_{21} = \begin{bmatrix} n_1 & 0 & n_2 & 0 & 0 & 0 \\ 0 & 0 & 0 & n_2 & 0 & n_1 \end{bmatrix},$$

$$\bar{\mathbf{n}}_{22} = \begin{bmatrix} 0 & 0 & n_1 & 0 & n_2 & 0 \\ 0 & n_2 & 0 & n_1 & 0 & 0 \end{bmatrix},$$

$$\bar{\mathbf{n}}_3 = \begin{bmatrix} n_1 & 0 & n_2 & 0 & 0 & 0 \\ 0 & n_2 & 0 & n_1 & 0 & 0 \\ 0 & 0 & n_1 & 0 & n_2 & 0 \\ 0 & 0 & 0 & n_2 & 0 & n_1 \end{bmatrix},$$

$$\bar{\mathbf{n}}_4 = \begin{bmatrix} n_1 & 0 & n_2 & 0 \\ 0 & n_2 & 0 & n_1 \end{bmatrix},$$

$$\bar{\mathbf{s}}_4 = \begin{bmatrix} n_2 & 0 & -n_1 & 0 \\ 0 & -n_1 & 0 & n_2 \end{bmatrix}.$$